\documentclass[a4paper,fleqn,usenatbib]{mnras}

\usepackage{newtxtext,newtxmath}
\usepackage[T1]{fontenc}
\usepackage{ae,aecompl}

\usepackage{graphicx}	% Including figure files
\usepackage{amsmath}	% Advanced maths commands
\usepackage{amssymb}	% Extra maths symbols
\usepackage{ulem}       %strikeout, added by WMY
\usepackage{color,xcolor} % textcolor, added by WMY

\title[Periodic mode changing in PSR J1048$-$5832]{Periodic mode changing
  in PSR J1048$-$5832}

\author[W. M. Yan et al.]{W. M. Yan,$^{1,2}$\thanks{E-mail: yanwm@xao.ac.cn (WMY)}
R. N. Manchester,$^{3}$
N. Wang,$^{1,2,4}$
Z. G. Wen,$^{1,2}$
J. P. Yuan,$^{1,2,4}$
\newauthor
K. J. Lee$^{5,6}$
and
J. L. Chen$^{7}$
\\
$^{1}$Xinjiang Astronomical Observatory, CAS, 150 Science 1-Street, Urumqi,
Xinjiang, 830011, China\\
$^{2}$Key Laboratory of Radio Astronomy, Chinese Academy of Sciences,
Nanjing 210008, China\\
$^{3}$CSIRO Astronomy and Space Science, Australia Telescope National
Facility, PO Box 76, Epping, NSW 1710, Australia\\
$^{4}$Xinjiang Key Laboratory of Radio Astrophysics, 150 Science 1-Street, Urumqi,
Xinjiang, 830011, China\\
$^{5}$Kavli Institute for Astronomy and Astrophysics, Peking University,
Beijing 100871, China\\
$^{6}$National Astronomical Observatories, Chinese Academy of Sciences,
Beijing 100012, China\\
$^{7}$Department of Physics \& Electronic Engineering, Yuncheng University,
044000, Yuncheng, Shanxi, China
}

\date{Accepted XXX. Received YYY; in original form ZZZ}

\pubyear{2019}

\begin{document}
\label{firstpage}
\pagerange{\pageref{firstpage}--\pageref{lastpage}}
\maketitle

\begin{abstract}
By analysing the data acquired from the 
Parkes 64-m radio telescope at 1369 MHz, we report on the phase-stationary non-drift
amplitude modulation observed in PSR J1048$-$5832. The high-sensitivity observations
revealed that the central and trailing components of the pulse profile of this
pulsar switch between a strong mode and a weak mode periodically. However, the leading
component remains unchanged. Polarization properties of the strong and weak modes are
investigated. Considering the similarity to mode changing, we argue that
the periodic amplitude modulation in PSR J1048$-$5832 is periodic mode changing. 
The fluctuation spectral analysis showed that the modulation period is very short
($\sim$2.1 s or $17 P_1$), where $P_1$ is the rotation period of the pulsar. We find
that this periodic amplitude modulation is hard to explain by existing models that
account for the periodic phenomena in pulsars like subpulse drifting. 

\end{abstract}


%%%%%%%%%%%%%%%%%%%%%%%%%%%%%%%%%%%%%%%%%%%%%%%%%%

%%%%%%%%%%%%%%%%% BODY OF PAPER %%%%%%%%%%%%%%%%%%

\section{Introduction}

For a given pulsar, even though the individual pulses vary in shape,
the mean pulse profile is highly stable with time. However, some
pulsars are observed to show emission variability over a wide range of
time-scales. It is well known that the Crab pulsar emits occasional
high-intensity nanosecond bursts that are called giant pulses
(e.g. \citealt{hkw+03}). High time resolution observations revealed
structure on timescales of a few microseconds in PSR J1136+1551
\citep{bart78} which is known as microstructure. The phenomenon of pulse
nulling was observed in a lot of pulsars, in which pulsed emission suddenly
turns off for several pulse periods and then just as suddenly turns on
\citep{rit76,ran86,big92a,viv95,wmj07,bjb+12,gjk12}. Mode changing is
another kind of emission change where pulsars switch between two or
more emission states (e.g. \citealt{wmj07}). The time-scale of mode
changing and nulling ranges from seconds to hours. Intermittent
pulsars switch between on and off emission states with cycles times
of days or even years \citep{klo+06,crc+12,llm+12,lsf+17}. 

Although most of these emission changes are largely considered to be random
processes, some changes are reported to show periodicities. Subpulse drifting is 
the best-known periodic emission variation in which subpulses drift in pulse
phase or longitude across a sequence of single pulses
(e.g. \citealt{wes06}). The subpulse drifting is a periodic pattern 
 with a characteristic spacing of the subpulses in pulse longitude ($P_2$) and pulse number
($P_3$). Periodic nulling is another common periodic change. 
Observations reveal that pulsar nulling is not always a stochastic process
\citep{rr09}. In some cases, the transitions between the null and the burst states
have shown periodicities \citep{rw07,rw08,hr07,hr09,rwb13,bmm17}. In recent years,
the periodic non-drift amplitude modulation, which looks very similar to
periodic nulling, has been reported in a number of pulsars \citep{bmm+16,mr17,ymw+19}.
It was reported that the periodic phase-stationary amplitude fluctuation is associated
with the presence of a core component in the pulse profile \citep{bmm+16,bmm+19}.

PSR J1048$-$5832 (B1046$-$58) is a young (20.3~kyr) pulsar, discovered
during a 1.5 GHz Parkes survey of the Galactic plane. It is located in
the Carina region at low Galactic latitude and has a spin period of
123.7~ms \citep{jlm+92}.  High-resolution X-ray observations by the
Chandra telescope revealed the existence of a faint asymmetric pulsar
wind nebula (PWN) surrounding this pulsar. However, no pulsed X-ray
emission has been detected from the pulsar \citep{gkp+06}.  By
analysing the EGRET data, \citet{klm+00} found possible $\gamma$-ray
pulsations at $E\ >$ 400~MeV and proposed that PSR J1048$-$5832 was
the counterpart of the unidentified $\gamma$-ray source 3EG
J1048$-$5840. More recently, $\gamma$-ray pulsations with a
double-peaked pulse profile \citep{aaa+09a} were clearly detected from
the pulsar for the first time at $E\ >$ 100~MeV by the Fermi
telescope. Radio observations show that the pulsar has a complex
profile and the profile varies significantly with frequency
\citep{kjm05,jkw06}. At 1.4~GHz, on average, the leading and central
components of the pulse profile exhibit a high degree of linear
polarization while the trailing component shows a very low fractional
linear polarization \citep{jk18}.

In this paper, we present the results of single-pulse observations at
1.4 GHz for PSR J1048$-$5832 that were made with the Parkes 64-m radio
telescope. We aim to explore previously unknown properties of the
periodic mode changing in PSR J1048$-$5832. Details of the
observations and data processing are given in
Section~\ref{sec:obs}. Properties of the observed periodic intensity
modulations are presented in Section~\ref{sec:results} and the
implications of the results are discussed in
Section~\ref{sec:discussion}. We conclude the paper in
Section~\ref{sec:conclusions}.

\section{Observations} \label{sec:obs}

The observational data used in our analyses were downloaded from the Parkes Pulsar Data
Archive which is publicly available online\footnote{\url{https://data.csiro.au}}
\citep{hmm+11}. The single-pulse observations were made using the Parkes 64-m radio
telescope on 2016 April 26 with the the H-OH receiver \citep{tgj90} and the
fourth-generation Parkes digital filterbank system PDFB4. See \citet{mhb+13} for further
details of the receiver and backend systems. For the observations reported in this paper,
the total bandwidth was 256 MHz centred at 1369 MHz with 512 channels across the band.
The data which were sampled every 256 $\mu$s last for 0.5 h and contain 14594 pulses.

The data were first reduced using the {\tt\string DSPSR} package \citep{vb11} to de-disperse
and produce single-pulse integrations which preserve information on individual pulses.
The pulsar's rotational ephemeris was taken from the ATNF Pulsar Catalogue
V1.60\footnote{\url{http://www.atnf.csiro.au/research/pulsar/psrcat/}} \citep{mhth05}. The
single-pulse integrations were recorded using the {\tt\string PSRFITS} data format
\citep{hvm04} with 256 phase bins per rotation period.  Strong narrow-band radio-frequency
interference (RFI) and broad-band impulsive RFI in the archive files were removed in affected
frequency channels and time sub-integrations, respectively. After RFI mitigation, the
single-pulse integrations were processed using the {\tt\string PSRCHIVE} pulsar data analysis
system \citep{hvm04}. The {\tt\string PSRSALSA} package \citep{wel16} which
  is freely available online\footnote{\url{https://github.com/weltevrede/psrsalsa}}, was used
  to carry out the analysis of fluctuation spectra.
Following \citet{ymv+11}, we carried out the flux density and polarization calibration for later
polarization analysis. The rotation measure (RM) value ($-155~\mathrm{rad~m^{-2}}$) was obtained
from \citet{qmlg95}. Stokes parameters were in accordance with the conventions outlined by
\citet{vmjr10}.

\section{Results} \label{sec:results}

Single-pulse properties of PSR J1048$-$5832 have not been published
before. We present and discuss the single-pulse properties for PSR J1048$-$5832 in this section.

\subsection{Nulling or mode changing?} \label{sec:mode}

\begin{figure}
\centering
\includegraphics[angle=0,width=\columnwidth]{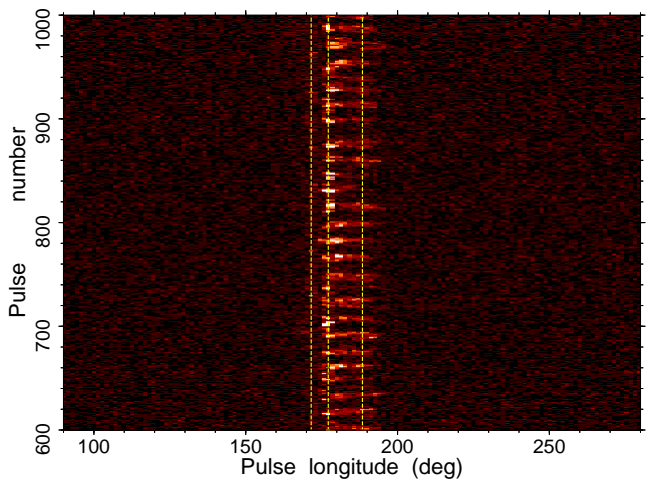}
\caption{A single-pulse stack of 400 successive pulses. The three vertical dashed lines
  show the longitudes of the leading, central and trailing pulse peaks of the mean pulse profile respectively.
  \label{fig:stack}}
     \end{figure}

\begin{figure}
\centering
\includegraphics[angle=0,width=\columnwidth]{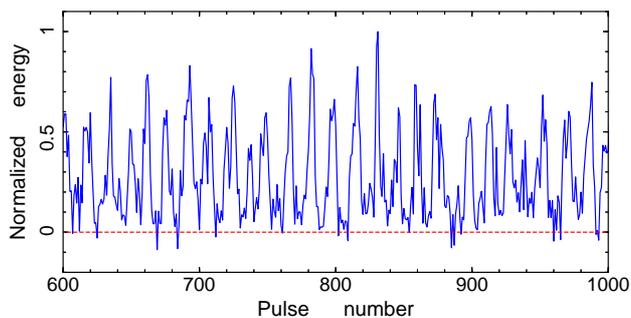}
\caption{Pulse energy variations for the pulse sequence shown in Fig.~\ref{fig:stack}.
        \label{fig:energy_var} }
\end{figure}

\begin{figure}
\centering
\includegraphics[angle=0,width=\columnwidth]{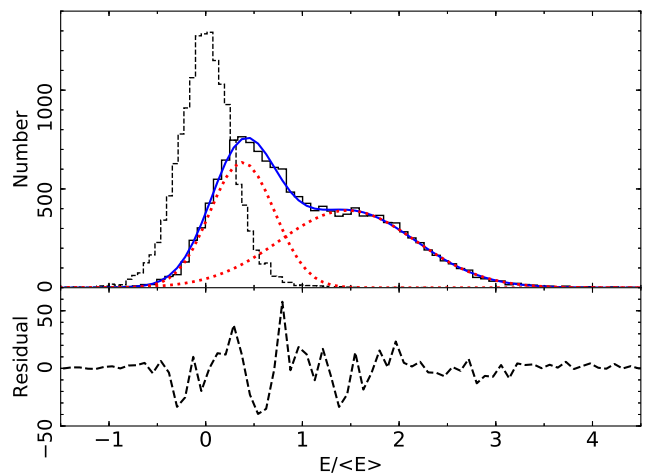}
\caption{Pulse energy distributions for the on-pulse (solid histogram) and off-pulse
  (dashed histogram) regions. The energies are normalised by the mean on-pulse energy.
  The blue solid line represents the fitting for the on-pulse energy distribution 
  based on the combination of two Gaussian components (red dotted lines).
  The lower panel shows the fit residuals.
  \label{fig:dist}}
\end{figure}

A pulse stack of 400 successive pulses is plotted in Fig.~\ref{fig:stack}. The pulse stack
clearly shows periodic modulation for the central and
trailing components, while the modulation in the leading component cannot be seen clearly.
Pulse energy variations for the same pulse sequence as in Fig.~\ref{fig:stack} are shown in
Fig.~\ref{fig:energy_var}. The pulse energy distributions for the on-pulse and off-pulse
windows for all pulses are presented in Fig.~\ref{fig:dist}. The 
on-pulse energy was determined for each individual pulse by summing the intensities of the pulse
phase bins within the on-pulse region. The on-pulse window was defined as the total longitude range
of the integrated pulse profile over which the pulse intensity significantly exceeds the baseline
noise (3$\sigma$). The off-pulse energy was calculated in the same way using an off-pulse window
of the same duration. In Fig.~\ref{fig:dist}, the off-pulse energy histogram shows a narrow
Gaussian distribution peaked about zero, while the broader on-pulse energy distribution shows two 
Gaussian-like components corresponding to a weak mode and a strong mode.

Fig.~\ref{fig:stack} gives the impression that the periodic modulation might be periodic nulling.
However, Fig.~\ref{fig:energy_var} and \ref{fig:dist} show that the observed periodic modulation
is rather different from pulse nulling in two key aspects.  Firstly, the pulse energy effectively
drops to zero in real null states, but the pulse energy in Fig.~\ref{fig:energy_var} is often larger
than zero even during the apparent ``null'' states. Secondly, 
real nulls have a similar energy distribution to the off-pulse noise.
Hence, a nulling pulsar tends to show a bimodal distribution in the on-pulse
energy histogram with two peaks, one at the zero energy and the other at the mean pulse energy.
Fig.~\ref{fig:dist} does show a bimodal on-pulse energy distribution with two peaks, but both peaks
are obviously larger than zero. This implies that the mean pulse energy in the apparent ``null''
states is larger than zero which conflicts with real null pulses. The polarization analysis in subsection
~\ref{sec:poln} shows that the central and trailing components of the pulse profile switch between a
stable strong mode and a stable weak mode periodically  while the leading component remains unaffected
by the modulation. This is very similar to mode changing. 
Therefore, in this paper, we argue that the observed periodic modulation in PSR J1048$-$5832 is
in fact periodic mode changing rather than periodic pulse nulling. 

\subsection{Fluctuation spectra} \label{sec:fluc}

\begin{figure}
\centering
\includegraphics[angle=-90,width=0.4\textwidth]{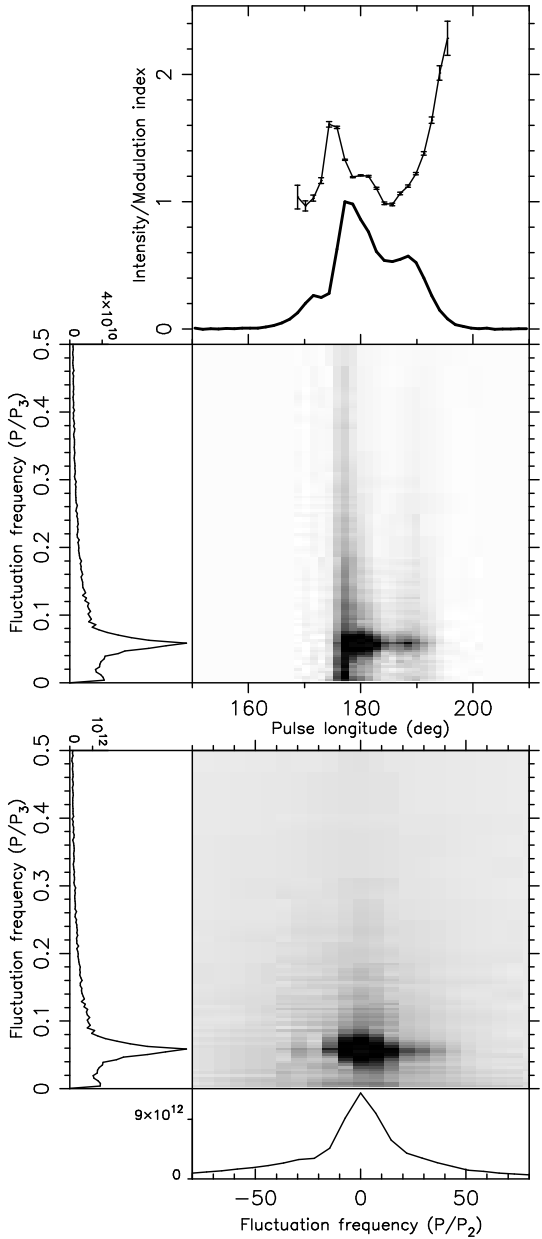}
\caption{Fluctuation analysis for PSR J1048$-$5832. The mean pulse 
         profile (solid line) and longitude-resolved modulation index 
         (points with error bars) are shown in the top panel. The LRFS with
         a side panel showing the horizontally integrated power is given 
         below this panel. The 2DFS with side panels showing horizontally (left)
         and vertically (bottom) integrated power is plotted below the LRFS.
         \label{fig:fluc}}
     \end{figure}

Fig.~\ref{fig:stack} clearly shows an apparently periodic modulation
for the central and trailing components.  To investigate the
periodicity further, we plotted the longitude-resolved modulation
index, the longitude-resolved fluctuation spectrum
\citep[LRFS,][]{bac70b} and the two-dimensional fluctuation spectrum
\citep[2DFS,][]{es02} in Fig.~\ref{fig:fluc}. The longitude-resolved
modulation index $m_i$ (points with error bars in the top panel) is a
measure of the amount of intensity variability from pulse to pulse for
each pulse longitude, defined by
\begin{equation}\label{eq:index}
m_i = \frac{\sigma_i}{\mu_i},
\end{equation} 
where $\mu_i$ and $\sigma_i$ are the average intensity and standard deviation at phase bin $i$
respectively.
The LRFS and 2DFS are effective for analysing subpulse modulation. The LRFS can be
used to determine which pulse phase bins present periodic subpulse modulation and with which
periodicities, while the 2DFS can be used to identify whether the subpulse drifts in pulse
longitude from pulse to pulse. The vertical frequency axis of both the LRFS and 2DFS 
corresponds to $P_1/P_3$, where $P_1$ is the rotation period of the pulsar and $P_3$ represents
the periodicity of intensity modulation that can be seen in the single-pulse stack
(Fig.~\ref{fig:stack}). The horizontal frequency axis of the 2DFS corresponds to $P_1/P_2$,
where $P_2$ represents the characteristic horizontal time separation between drifting bands.
For more details regarding the techniques of analysis, please refer to \citet{wes06}.

From Fig.~\ref{fig:fluc}, we can see that the leading component and the saddle region
between the central and trailing components have the lowest modulation index. However, the average
intensity is much higher for the saddle region. From the definition of modulation index
(Equation~(\ref{eq:index})), there is just a very weak modulation for the leading component.
The side panel of the LRFS shows a 
clear spectral peak occurring between 0.0571 and 0.0581 cycles per period (cpp). This spectral
feature corresponds to the characteristic $P_3$ of $17.4\pm 0.2\;P_1$ for the periodic modulation.
In the LRFS, the spectral feature is stronger in the central and trailing components,
implying that most of the periodic modulation power is associated with these
components. In addition to the well-defined spectral feature, there are white noise components
which are clearly seen as two vertical darker bands in the LRFS around the leading edge of the
central component and the peak of the trailing component respectively. These indicate
non-periodic stochastic fluctuations at these pulse longitudes. The 2DFS is basically 
symmetric about the vertical axis, so subpulses in successive pulses do not drift on average to
later or earlier pulse longitudes.

\begin{figure}
\centering
\includegraphics[angle=-90,width=0.4\textwidth]{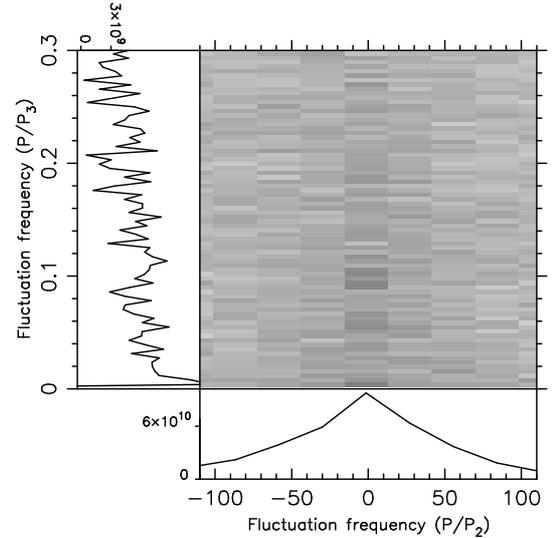}
\caption{The 2DFS for the leading component. See Fig.~\ref{fig:fluc} for further details.
         \label{fig:2dfs_leading}}
     \end{figure}

Furthermore, to make it clear whether there is any low level periodic behavior in the leading 
  component hidden by the high signal to noise variations in the other two components, we plotted the
  2DFS only for the leading component in Fig.~\ref{fig:2dfs_leading}. A red-noise-like component is
  strongest below $P_1/P_3$ = 0.005 cpp, corresponding to fluctuation on a timescale of 200 pulse periods
  and above. There is no visible periodic feature in the leading component around $P_3\simeq17.4P_1$.
  
\subsection{Mode separation} \label{sec:separation}

To obtain the relative abundances of the strong and weak modes, modeling of the on-pulse energy
histogram was carried out by fitting two Gaussian functions. 
  We used the {\tt\string curve\_fit} function of the {\tt\string scipy} python module, 
  which uses non-linear least squares to fit a function to data, to fit the sum of two Gaussian functions
  to the histogram and the parameters of the fitted Gaussians are listed in Table~\ref{tab:gaus}.
The results are presented in Fig.~\ref{fig:dist}
  which shows that the two modes overlap with no clear boundary between them. Then the relative abundances of each
  mode can be simply obtained by calculating the area under
  each Gaussian function. The results show that this pulsar spends 43\% in the weak mode and 57\% in the strong mode.

 \begin{table}
\begin{center}
\centering
\begin{minipage}[]{120mm}
\caption{Parameters of fitted Gaussian functions.
\label{tab:gaus}}
\end{minipage}
\begin{tabular}{lccc}
\hline
 Mode &Amplitude  &Mean  & Standard Deviation  \\
\hline
Weak mode    &634   &0.38   &0.33     \\
Strong mode   &391   &1.47   &0.71    \\
\hline
\end{tabular}
\end{center}
\end{table} 

\begin{figure*}
\centering
\includegraphics[angle=0,width=0.9\textwidth]{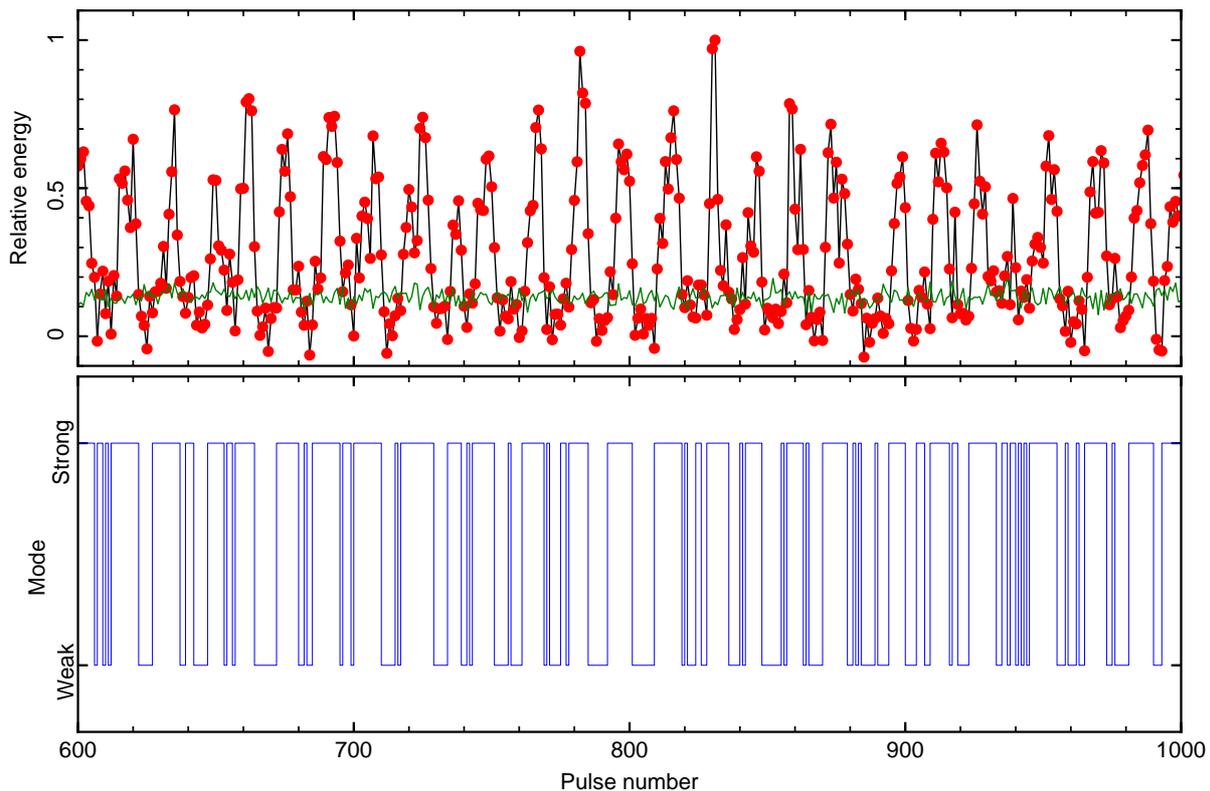}
\caption{The same pulse energy sequence as Fig.~\ref{fig:energy_var} (upper
         panel) with the corresponding identified strong/weak mode
         (lower panel). In the upper panel, a red dot is the on-pulse energy
         of a given pulse and the green line represents the three times level of
         the uncertainty in the pulse energy estimate for individual pulses
         (see text for details). Pulses for which pulse
         energy below the green line were tagged as weak-mode pulses and pulses
        above the green line were tagged as strong-mode pulses.
\label{fig:mode} }
\end{figure*}

Separation of the strong and weak modes is required for more detailed
investigation. However, the above mentioned Gaussian fitting method cannot distinguish pulses in the
overlapped area of the two Gaussian functions.
Following \citet{bgg10}, we identified the two modes by
comparing the on-pulse energy of individual pulses with the system
noise level.  The uncertainty in the on-pulse energy $\sigma_{\rm on}$
is given by $\sqrt{N_{\rm on}}\sigma_{\rm off}$ , where $N_{\rm on}$
is the number of on-pulse phase bins, estimated from the mean
pulse profile, and $\sigma_{\rm off}$ is the rms of the off-pulse
region for individual pulses. Pulses with on-pulse energy smaller than
$3\sigma_{\rm on}$ were classified as weak-mode pulses and the others
were classified as strong-mode pulses.  Fig.~\ref{fig:mode} shows the
results of the state separation for the same pulse energy sequence as
Fig.~\ref{fig:energy_var}.  By performing the state separation in this
way for all data, we found that 65\% of the time for PSR J1048$-$5832
was in the strong mode and 35\% was in the weak mode.
There are some differences in values of the relative abundances of the strong and weak modes obtained
  using two methods. This is because neither method can completely separate the two modes. In the absence of
  perfect mode separation, we acknowledge that there is always a possibility of mode mixing which may affect the
  conclusions in our subsequent analysis for profile shape evolution, polarization, etc.

\subsection{Polarization} \label{sec:poln}

\begin{figure*}
\centering
\includegraphics[angle=0,width=0.9\textwidth]{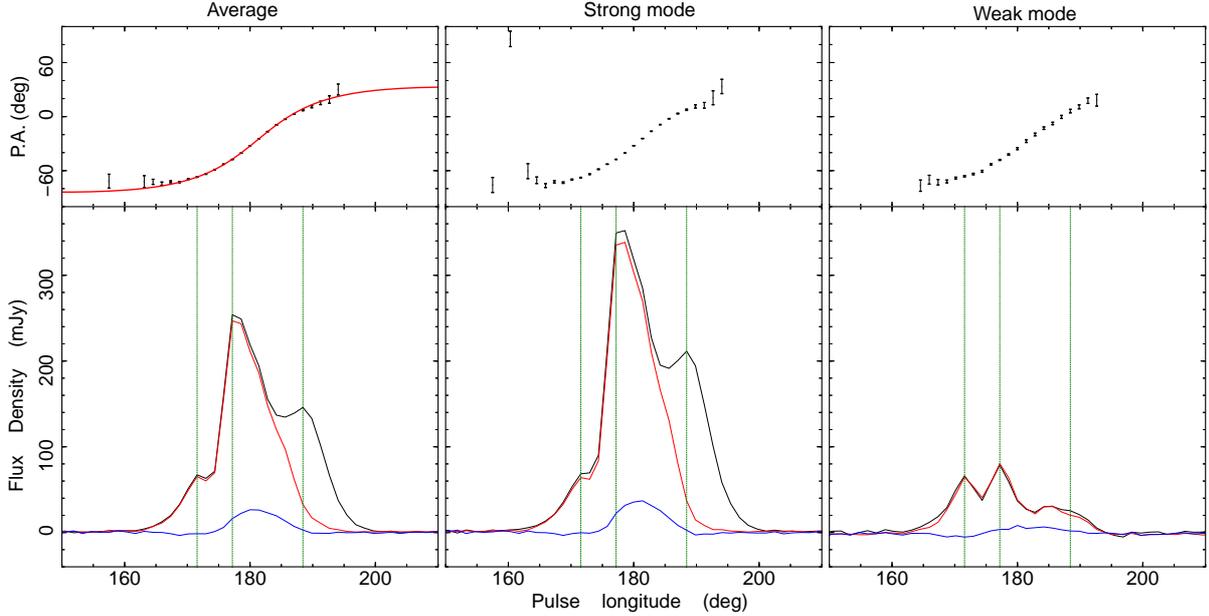}
\caption{Polarization profiles for the average of all pulses (left), the strong-mode pulses (middle),
  and the weak-mode pulses (right). The lower panels show the pulse profiles for total intensity
  (black line), linearly polarized intensity (red line), and circularly polarized intensity
  (blue line). The upper panels give the position angles of the linearly polarized emission.
  The three vertical lines show the longitudes of the leading, central and trailing pulse
  peaks of the mean pulse profile for all pulses respectively. The S-shaped solid line in the
  top left panel represents the best fit of the PA swing.
  \label{fig:poln} }
\end{figure*}

To compare the weak mode and the strong mode further, we analysed the polarization properties for
both modes. We show mode-separated polarization profiles in Fig.~\ref{fig:poln}, and
Table~\ref{tab:poln} gives a summary of the flux density and polarization parameters for different
modes. Columns (2) to (4) give the peak flux density for the leading ($S_{\rm pk,leading}$), central 
($S_{\rm pk,central}$) and trailing ($S_{\rm pk,trailing}$) components respectively.
The next column gives the mean flux density $S$. Pulse width at 50 per cent of the peak
flux density $W_{50}$ is given next. The next three columns give the fractional linear polarization
$\langle L \rangle/S$, the fractional net circular polarization $\langle V \rangle /S$ and the
fractional absolute circular polarization $\langle |V| \rangle/S$, where $\langle \rangle$ means
an average across the pulse profile. Polarization properties of each profile component in different
  modes are given in Table~\ref{tab:comp}. We defined the outer boundaries of the
  leading and trailing components as the first point and the last point where the pulse intensity significantly
  exceeds the baseline noise (3$\sigma$) respectively, 
  while the inner boundaries were defined as the points of minimal intensity between adjacent components.
  The width of a given component was measured as the longitude range between its two boundaries. The second
  column $\phi_{\mathrm{pk}}$ in each mode in Table~\ref{tab:comp} represents the longitudinal position of
  component peaks.

\begin{figure}
\centering
\includegraphics[angle=0,width=0.4\textwidth]{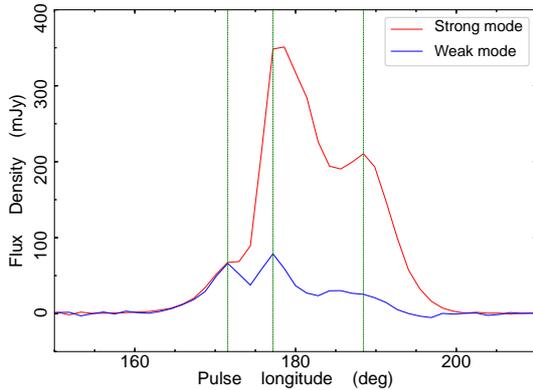}
\caption{The comparison of the average profile for pulses in the strong mode and the weak
  mode. The three vertical lines show the longitudes of the leading, central and trailing pulse
  peaks of the mean pulse profile for all pulses respectively. \label{fig:comp}
}
\end{figure}

In Fig.~\ref{fig:poln}, the polarization profile for the average
profile in the left panel is in good agreement with previously
published results \citep{rwj15,jk18}. The pulse profile has multiple
overlapping features. The leading and the central parts of the profile
have very high fractional linear polarization, whereas the trailing
part of the profile is essentially unpolarized. Except for the central
component, circular polarization is quite weak. The position angle
(PA) swing is S-shaped through the profile and relatively shallow in
the leading and trailing components. This means that neither the
leading component nor the trailing component can be core
emission. Based on a rotating vector model \citep[RVM,][]{rc69a} fit
to the PA variations and other
constraints, \citet{rwj15} gave an estimate of the impact parameter
(the closest approach of the line of sight to the magnetic axis)
  $0\degr<\beta<7\fdg5$ and suggested that magnetic axis was
  relatively aligned with the rotation axis, with an inclination angle
  $\alpha<50\degr$. With the constraints on $\alpha$ and $\beta$ reported
    by \citet{rwj15}, we carried out an RVM fit to the PA swing for the average profile
    using the {\tt\string ppolFit} program of {\tt\string PSRSALSA}. The S-shaped solid curve in
    the top left panel in Fig.~\ref{fig:poln} shows the best fit to the PA with $\alpha=4\fdg1$ and 
    $\beta=0\fdg7$. The best fit has a reduced $\chi^{2}=5.4$ . The impact parameter $\beta$ is
    quite small implying that the central component is probably the core component.
  The strong mode has very similar polarization
properties, which just shows stronger central and trailing components.
However, the weak mode shows a very different polarization
profile. The leading component is the weakest in the strong mode
whereas the trailing component is the weakest in the weak mode. The
central component in the weak mode is also relatively weak. The
central component remains nearly 100 per cent linearly polarized in
different modes. The trailing component is essentially unpolarized in
the strong mode, while it becomes 100 per cent linearly polarized in
the weak mode.  The fractional circular polarization in the weak mode
is relatively high with a sense reversal between the leading and
central components.  Both the PA swing and the overall pulse width are
identical for the two modes and such 
  behavior has also been seen in PSRs J1822$-$2256 \citep{bm18} and 
  J2006$-$0807 \citep{bpm19},  which suggests that the emission of the
two modes comes from the same general emission region. Additionally, we
compare total intensity of the two modes in Fig.~\ref{fig:comp}. This
figure clearly shows that the total intensity of the leading component
remains unchanged in the two modes. That is, the main difference
between the two modes lies in the central and trailing
components. This implies that the observed periodic mode changing
arises from variations in the central and trailing components while
the leading component is not affected by the mode changing.

\begin{table*}
\begin{center}
\centering
\begin{minipage}[]{120mm}
\caption{Flux density and polarization parameters for the strong and weak modes.}
\label{tab:poln}
\end{minipage}
\begin{tabular}{lcccccccc}
\hline
  Mode &$S_{\rm pk,leading}$ &$S_{\rm pk,central}$ &$S_{\rm pk,trailing}$ &$S$ & $W_{50}$
  &$\left.\left\langle{L}\right\rangle\middle/S\right.$
  &$\left.\left\langle{V}\right\rangle\middle/S\right.$
  &$\left.\left\langle\left|{V}\right|\right\rangle\middle/S\right.$\\
  & (mJy) & (mJy) & (mJy) & (mJy)  &($\degr$)  &(\%)  &(\%)  &(\%)  \\
\hline
Average  &66  &253  &145  &10.1  &14.9  &74.6  &6.2  &7.2     \\
Strong mode &67  &348  &211  &16.1  &15.1  &69.3  &5.8  &6.7   \\
Weak mode   &66  &79   &25   &3.9   &10.4  &95.6  &2.0  &10.8   \\
\hline
\end{tabular}
\end{center}
\end{table*}

\begin{table*}
\begin{center}
\centering
\begin{minipage}[]{140mm}
\caption{Polarization parameters of each pulse profile component for the strong and weak modes.}
\label{tab:comp}
\end{minipage}
\begin{tabular}{cccccccccccc}
\hline
Component  &\multicolumn{5}{c}{Strong mode} &  & \multicolumn{5}{c}{Weak mode}\\
\cline{2-6}
\cline{8-12}
  &  Width & $\phi_{\mathrm{pk}}$  &$\left.\left\langle{L}\right\rangle\middle/S\right.$
  &$\left.\left\langle{V}\right\rangle\middle/S\right.$
  &$\left.\left\langle\left|{V}\right|\right\rangle\middle/S\right.$  &  &Width & $\phi_{\mathrm{pk}}$  &$\left.\left\langle{L}\right\rangle\middle/S\right.$ &$\left.\left\langle{V}\right\rangle\middle/S\right.$ &$\left.\left\langle\left|{V}\right|\right\rangle\middle/S\right.$ \\
  & ($\degr$)   &($\degr$)  &(\%)  &(\%)  &(\%) &  & ($\degr$)   &($\degr$)  &(\%)  &(\%)  &(\%) \\
\hline
Leading component      &9.9   &172.2  &91.9   &$-2.0$ &3.3  &  &11.3  &172.2  &92.3   &$-9.7$ &9.7 \\
Central component      &12.7  &179.3  &91.8   &9.1    &9.1  &  &8.5   &177.9  &100.0  &8.3    &9.0 \\
Trailing component     &14.1  &189.2  &24.2   &1.7    &3.1  &  &16.9  &186.4  &90.0   &9.0    &17.9 \\
\hline
\end{tabular}
\end{center}
\end{table*}

\section{Discussion} \label{sec:discussion}

More and more periodic amplitude fluctuation phenomena have been observed in radio pulsar emission. 
\citet{bmm+16} carried out a detailed fluctuation spectral analysis for 123 pulsars that were
observed at 333 and/or 618 MHz by the GMRT telescope. They revealed periodic features in 57 pulsars, 
including 29 pulsars where the periodic features show no phase variation but periodic amplitude
fluctuation. They discovered the dependence of the drifting phenomenon on the spin-down luminosity
$\dot{E}$. In those 57 pulsars, subpulse drifting is only seen in pulsars with
$\dot{E}< 5\times$10$^{32}\ {\rm erg\,s^{-1}}$ and the drifting periodicity $P_3$ is anticorrelated
with $\dot{E}$, whereas other pulsars with $\dot{E}>5\times$10$^{32}\ {\rm erg\,s^{-1}}$ all showed
phase-stationary non-drift amplitude modulation. \citet{bmm+19} verified this dependence using a
larger sample of drifting measurements. PSR J1048$-$5832 has an
$\dot{E}=2\times$10$^{36}\ {\rm erg\,s^{-1}}$ \citep{wmp+00}, and thus lies in the non-drift amplitude
modulation group.

Based on empirical classification for the pulse profiles of radio pulsars, \citet{ran83,ran90,ran93}
suggested a core/cone model for the radio emission beam. In this model, the emission beam consists
of multiple nested cones around a central core component. Different numbers of pulse components
result from different line of sight (LOS) trajectories across the beam. \citet{bmm+16} found that
the periodic non-drift amplitude modulation was usually found in pulsars with core components.
By analysing phase variations corresponding to the drifting features, \citet{bmm+19} proposed a
classification scheme for the phenomenon of subpulse drifting, and they argued that the drifting
classification was associated with the profile types and LOS geometry. The periodic non-drift
amplitude modulation corresponds to an LOS close to the beam center that sweeps across both the core
and the cone regions. In this case, pulsars show the presence of a central core component in their
profiles where the drifting is absent, while the conal components show relatively smaller phase
variations. However, PSR J1048$-$5832 does not match the pattern. Polarization profiles suggest that
the leading and trailing components may be conal components  (see Fig.~\ref{fig:poln}) and the small
impact parameter implies that the central component is probably the core component. But the fluctuation
spectra shown in Fig.~\ref{fig:fluc} show that the central component and the trailing conal component
exhibit strong periodic modulation, while the leading conal component is unmodulated, which is
not consistent with the simple LOS geometry model of \citet{bmm+19}. 

Periodic nulling has been reported in some pulsars 
\citep{hr07,hr09,rw08,rwb13,gjw14,gyy+17,bmm17,bm18,bpm19,bm19}. Observed
periodicities in the amplitude modulation of some components of some
pulsars are very similar to periodic nulling, and therefore
\citet{bmm17} suggested that they have a common origin. This supports
the argument of \citet{wmj07} that nulling and mode changing are
different manifestations of the same phenomenon. Here we therefore
suggest that periodic nulling is the extreme form of periodic mode
changing and that the two phenomena are just different categories of
the same physical phenomenon.  Some investigators argue that periodic
nulling can be explained by the rotating subbeam carousel model
\citep{hr07}, in which the evenly spaced conal subbeams rotate around
a central core component. Empty LOS cuts or extinguished subbeams can
repeat after certain spin periods of the pulsar, thus giving rise to
periodic nulling. However, \citet{bmm17} found that the missing LOS
model did not adequately explain the periodic nulling observed in
pulsars with core components. Additionally, \citep{gyy+17} found that
the missing LOS model is probably not applicable to the
quasi-periodic changes between the null and burst states seen in PSR
J1741$-$0840.  For the periodic amplitude modulation of PSR
J1048$-5832$ reported in this paper, the probable central core
component and the trailing conal component switch between a strong
mode and a weak mode periodically. The weak mode shows relatively weak
emission, but not complete nulls. This does not fit very well with the
missing LOS model. Furthermore, the leading conal component is not
modulated, which cannot be explained with the rotating subbeam
carousel model.  The polarization difference of the trailing component
in different modes is also not readily explained with the carousel
model.

\citet{mr17} reported a similar periodic non-drift modulation in PSR
B1946+35, in which the highly periodic amplitude modulation was seen
in the core component. This could also not be explained using the
carousel model and was considered as an entirely new phenomenon.
In a recent work by \citet{bm19}, the periodic amplitude modulation
  has also been seen in the core single pulsar PSR J0826+2637.
A periodic phase-stationary intensity
modulation occurring in the interpulse and the main pulse was observed
in PSR J1825$-$0935 \citep{fw82,gjk+94,bmr10,lmr12,hkh+17,ymw+19}. Similar
to PSR J1048$-5832$, the main pulse of
PSR J1825$-$0935 is partially modulated with an unmodulated trailing
component. The periodic amplitude modulation in PSRs J1048$-$58321,
B1946+35, J0826+2637 and J1825$-$0935 is probably regular switching between two
emission states, i.e., mode changing. We argue that
this amplitude modulation is periodic mode changing which results from
similarly periodic fluctuations in the magnetospheric field currents
that are not directly related to subpulse drifting.

\section{Conclusions} \label{sec:conclusions}

We report here a periodic phase-stationary non-drift amplitude modulation observed in PSR J1048$-$5832
using the Parkes 64-m radio telescope. As argued above, the amplitude modulation cannot be reconciled
with the subpulse drifting phenomenon in any existing geometrical context. It seems
to require emission-state changing in the magnetospheric field currents with the same periodicity. This
is similar to the mode changing phenomenon, though the time-scale of mode changing is usually longer.
Therefore we suggest the periodic amplitude modulation in PSR J1048$-$5832 is periodic mode changing.
Discovering and observing more samples with this amplitude modulation using large radio telescopes such
as FAST will throw more light on the nature of this phenomenon.

\section*{Acknowledgements}

This work is supported by National Basic Research Program of China
(973 Program 2015CB857100), National Natural Science Foundation of China
(Nos. U1831102, U1731238, U1631106, 11873080, U1838109),
the Strategic Priority Research Program of Chinese Academy of Sciences
(No. XDB23010200), the National Key Research and Development Program of China
(No. 2016YFA0400800), the 2016 Project of Xinjiang Uygur Autonomous Region of
China for Flexibly Fetching in Upscale Talents and the
CAS ``Light of West China'' Program (2018-XBQNXZ-B-023, 2016-QNXZ-B-24 and
2017-XBQNXZ-B-022). We thank R. Yuen for valuable discussion.
The Parkes radio telescope is
part of the Australia Telescope, which is funded by the Commonwealth of
Australia for operation as a National Facility managed by the Commonwealth
Scientific and Industrial Research Organisation.

%%%%%%%%%%%%%%%%%%%%%%%%%%%%%%%%%%%%%%%%%%%%%%%%%%



%\bibliographystyle{mnras}
%\bibliography{journals,modrefs,psrrefs,crossrefs}

%%%%%%%%%%%%%%%%%%%%%%%%%%%%%%%%%%%%%%%%%%%%%%%%%%

% Don't change these lines
\bsp	% typesetting comment
\label{lastpage}
\end{document}